\documentclass{elsart}

\def\ie{{\it i.e.,~}}
\def\lesssim{\raise0.3ex\hbox{$<$}\kern-0.75em{\lower0.65ex\hbox{$\sim$}}}
\def\gtrsim{\raise0.3ex\hbox{$>$}\kern-0.75em{\lower0.65ex\hbox{$\sim$}}}

    \def\independenT#1#2{\mathrel{\setbox0\hbox{$#1#2$}%
    \copy0\kern-\wd0\mkern4mu\box0}} 

\usepackage{epsfig}

\begin{document}
\begin{frontmatter}
\title{A proposal for testing subcritical vacuum pair production with high 
power lasers}

\author[OU,RAL]{G. Gregori}
\author[Wroclaw,DUB]{D. B. Blaschke}
\author[RAL]{P. P. Rajeev}
\author[LLNL]{H. Chen}
\author[RAL]{R. J. Clarke}
\author[OU]{T. Huffman}
\author[OU]{C. D. Murphy}
\author[Saratov]{A. V. Prozorkevich}
\author[ANL]{C. D. Roberts}
\author[Rostock]{G. R\"opke}
\author[DU,J]{S. M. Schmidt}
\author[Saratov]{S. A. Smolyansky}
\author[LLNL]{S. Wilks}
\author[RAL]{R. Bingham}

\address[OU]{Department of Physics, University of Oxford, Parks Road, Oxford, 
OX1 3PU, UK}
\address[RAL]{Rutherford Appleton Laboratory, Chilton, Didcot OX11 0QX, UK}
\address[Wroclaw]{Institute for Theoretical Physics, University of Wroclaw, 
50-204 Wroclaw, Poland}
\address[DUB]{Bogoliubov Laboratory for Theoretical Physics, Joint Institute 
for Nuclear Research, RU-141980, Dubna, Russia}
\address[LLNL]{Lawrence Livermore National Laboratory, 7000 East Avenue, 
Livermore, CA 94550, USA}
\address[Saratov]{Saratov State University, RU-410026, Saratov, Russia}
\address[ANL]{Physics Division, Argonne National Laboratory, Argonne, 
IL 60439-4843, USA}
\address[Rostock]{Institut f{\"u}r Physik, Universit{\"a}t Rostock, 
Universit{\"a}tsplatz 3, 18051 Rostock, Germany}
\address[DU]{Technische Universit\"at Dortmund, Fakult\"at Physik \& DELTA, 
44221 Dortmund, Germany}
\address[J]{Forschungszentrum J\"ulich GmbH, 52428 J\"ulich, Germany}

\date{\today}

\begin{abstract}
We present a proposal for testing the prediction of non-equilibrium quantum 
field theory below the Schwinger limit. The proposed experiments should be 
able to detect
a measurable number of gamma rays resulting from the annihilation of pairs in 
the focal spot of two opposing high intensity laser beams. We discuss the 
dependence of the
expected number of gamma rays with the laser parameters and compare with the 
estimated background level of gamma hits for realistic laser conditions.
\end{abstract}

\begin{keyword}
quantum electrodynamics\sep%
non-pertubative limit\sep%
pair production\sep%
high intensity lasers.%

\end{keyword}

\end{frontmatter}

\section{Introduction}
Current generation of high-repetition and high-intensity lasers as well as 
$4^{\rm th}$ generation light sources have opened the possibility to
experimentally 
test quantum electrodynamics (QED) at the level where the details of multiple 
order expansions in the field propagators can be verified with measurable 
observables. 
This {\it subcritical} field limit is of extreme scientific interest since it 
will allow progress of the physics research into completely new realms with 
the generation of novel and unexplored states of matter: electron-positron 
plasmas and {\it excited} vacuum states. 
Moreover, measured deviations from the predicted QED processes could indicate 
correction from quantum gravity or Lorentz-violations. 
These experiments can provide the required benchmark for cosmological vacuum 
particle production between the Planck and the GUT (Grand Unified Theory) era. 
In such cases the external field provided by the laser is replaced by the 
interaction of a massive scalar field with the background space-time, but the 
governing equations of non-equilibrium quantum field theory (NeqQFT) still 
hold in the same form. 

In this paper we will discuss the theoretical framework implemented to 
calculate pair production at the subcritical limit and we will solve the 
governing NeqQFT equations in some idealized, yet representative, case 
describing the interaction of two high intensity laser beams in vacuum.
In the second part of this paper, we will investigate the application of these 
approaches to a realistic experimental setup. In particular, we will discuss 
the possibility of observing gamma rays from pair annihilation in the laser 
focal spot and we will compare this number with the estimated background level.
We will use as examples for the experimental configuration both the Gemini 
laser system, recently commissioned at the Rutherford Appleton Laboratory (UK),
and the current generation of petawatt lasers such as the Vulcan laser at the 
Rutherford Appleton Laboratory. 
We will also compare optical pair production versus x-ray Free Electron Laser 
(FEL) sources and discusses the differences among those approaches.

\section{Subcritical pair production}
Due to its intrinsic simplicity, vacuum pair production is the proposed 
experiment of choice to test QED in the subcritical field limit. 
It is well known from basic QED that high energy photon-photon scattering can 
result in the production of electron-positron pairs. 
This process and the corresponding reverse interaction of electron-positron 
annihilation play an important role in determining the overall opacity of the 
interstellar and intergalactic medium which in turn relates to the correct 
estimates of the intrinsic luminosity of stellar objects. 
At the same time, pair production provides a mean by which large fluxes of 
positrons could be generated in the vicinity of active galactic nuclei. 
One of the most exciting results of $\gamma$-ray astronomy has indeed been the 
detection of the 0.511 MeV emission line from the Galactic center \cite{p1}. 
Massive compact and quasi stellar objects are also source of intense beams of 
optical and infrared radiation and high-order low-energy multiphoton 
interactions which result in pair production are also possible. 
Moreover, processes involving a massive neutral scalar field in a dynamical 
background (either due to an external semi-classical field or a space-time 
metric) are applicable to cosmological problems such as vacuum particle 
production at the Planck time or reheating after GUT scale inflationary 
expansion \cite{p2,p3,p4}. 
Another process closely related, and described within the same NeqQFT 
framework, is the thermal radiation arising from particle production near the 
event horizon of a black hole, commonly known as Hawking effects \cite{p5}, as 
well as the Unruh effect \cite{p6}, which is seen in uniformly accelerated 
detectors at relativistic speed. Recently, suggestions have been brought 
forward that the Unruh effect could be detected with the currently available 
Gemini laser at the Rutherford Appleton Laboratory \cite{p7}. 

On the theoretical side, subcritical vacuum breakdown is non-perturbative. 
Solutions exist only for idealized configurations and experimental verification
is important for the correct understanding of the process and its relevance to 
the total interstellar opacity. 
At the same time this work could provide the first high-density 
electron-positron plasma to test and simulate a variety of astrophysical 
environments. 
With the advent of chirped pulse amplification (CPA) techniques \cite{p8} and 
progresses in x-ray free electron lasers (FELs) \cite{p9} it now becomes 
possible to generate very large numbers of coherent photons (i.e., high 
electric fields) in both the optical and the x-ray wavelengths. 
For any astrophysical object we define ?compactness? as the ratio between the 
total heating divided by its physical size \cite{p10}. 
In our context we are often dealing with high compactness objects, where pairs 
are primarily created by photon-photon collisions, and the energy loss is 
negligible as all the pairs remain confined within the laser focal spot. 
For the production of an electron-positron pair, the center of mass energy of 
the two photons must exceed $2mc^2$, which precludes the creation of the pair 
by the collision of two single optical or x-ray photons. 
In strong electromagnetic fields, however, the interaction is not limited to 
initial states with two single photons, but allows multiphoton processes 
\cite{p11}
\begin{equation}
	N_1 (\hbar \omega_1) + N_2 (\hbar \omega_2) \rightarrow e^+ + e^-,
\end{equation}
where $N_1$ and $N_2$ are large integers.
Experimental verification of the collision between $\sim$4 coherent optical 
photons with one gamma ray photon with energy $\sim$30 GeV 
(created by Compton backscatter of another optical photon against an high 
energy electron beam) and the corresponding production of electron positron 
pairs has been demonstrated at the SLAC facility \cite{p12,c1}. 
Here, instead, we are interested in the more extreme case of a vacuum breakdown
driven by large number of low energy photons $N_1\sim N_2 = N \gg 1$ and 
$\omega_1 \sim \omega_2=\omega \ll m$ (we use, as customary, natural units 
where $\hbar=c=1$). 
This is an example of NeqQFT where quantum mean field approaches have been 
proposed \cite{p2,p13,p14} but need experimental validation. 
The basic of these approaches is the so-called quantum Vlasov equation. 
In spinor QED assuming an external semi-classical electric field, it is 
possible to show that, for fermions, the particle number operator satisfies an 
equation of the type \cite{p2,p14}
\begin{eqnarray}
\frac{d f_{\bf k}(t)}{dt} &=&
\frac{\dot{\Omega}_{\bf k}(t) \epsilon_{\perp}(t)}
{2 \Omega_{\bf k}(t) \epsilon_{\|}(t)} 
\nonumber \\
&& \times \int_{-\infty}^t du 
\left\{ \frac{\dot{\Omega}_{\bf k}(u) \epsilon_{\perp}(u)}
{\Omega_{\bf k}(u) \epsilon_{\|}(u)}
\left[1-2 f_{\bf k}(u)\right] 
\cos \left[2 \int_u^t d\tau \Omega_{\bf k}(\tau)\right] \right\},
\label{qve}
\end{eqnarray}
which is known as a quantum Vlasov equation, and
\begin{equation}
\Omega_{\bf k}^2 = m^2+k_{\perp}^2+(k_{\|}-eA)^2,
\end{equation}
with
\begin{equation}
\epsilon_{\perp}^2 = m^2+k_{\perp}^2,
\end{equation}
\begin{equation}
\epsilon_{\|}^2 = (k_{\|}-eA)^2,
\end{equation}
and $k_{\perp}$ ($k_{\|}$) is the momentum perpendicular (parallel) to the 
linearly polarized electric field ${\bf \cal E} = -{\bf \dot{A}}$ (in the 
Coulomb gauge).
The total electron-positron number per unit volume is then obtained by 
integrating over all the momenta,
\begin{equation}
	{\cal N}(t) = 2 \int \frac{d^3 k}{(2\pi)^3} f_{\bf k}(t).
	\label{qk}
\end{equation}

The quantum Vlasov equation has a non-Markovian character given by the factor 
$[1-2 f_{\bf k}(t)]$ arising from quantum statistics as it takes into account 
the full history of the distribution function. 
It simply says that the pair production rate will be affected by the particles 
already present in the system. 
However, in the case of weak (subcritical) fields, such an effect can be often 
neglected \cite{p16}. 
The non-Markovian character is also inherent in the phase oscillations 
represented by the cosine term. 
This is related to quantum coherence, resulting from the fact that when the 
two pairs are created, they are initially fully correlated (\ie entangled). 
The time-scale for these quantum coherence effects to wash out is in the order 
of $\tau_{qu} \sim 2\pi/\Omega_{\bf k} \sim 2\pi/m$ \cite{p2,p13}.
In order for the statistical description of pair production to be valid, this 
time must be shorter than the time required to produce the pairs.
This, for small $k$, can be estimated from Eq.~(\ref{qve}) to be \cite{p13}
\begin{equation}
\tau_{cl} \sim \left[\frac{\dot{\Omega}_{\bf k}(t) \epsilon_{\perp}(t)}
{\Omega_{\bf k}(t) \epsilon_{\|}(t)}\right]^{-1} \sim \frac{m}{e {\cal E}}.
\end{equation}
In the semi-classical case we also need to assume that the external field 
remains approximately constant during particle generation, that is
$\tau_{cl} < \tau_{pl}$ \cite{p13}, where $\tau_{pl}$ is the characteristic 
time associated to collective plasma fluctuations: 
$\tau_{pl}=2 \pi/\omega_{pl} = 2 \pi (m/e^2 n_{av})^{1/2}$ with $n_{av}$ 
the average pair density.

Equation (\ref{qk}) is just a formal solution, and a few words are necessary 
in order to correctly interpret its meaning. 
This point has been discussed in the literature \cite{p2,p3}, and it stems from
the fact that the number of pairs does not commute with the Hamiltonian 
(indeed it is not a constant of motion). 
This follows directly from the uncertainty principle. 
If we have $N_{ep}$ pairs, the uncertainty relation reads as
\begin{equation}
	\Delta E \Delta t = \Delta (N_{ep} m) \Delta t \sim 1,
\end{equation}
and the uncertainty in the particle number is $\Delta N_{ep}\sim 1/m \Delta t$.
This implies that the particle number is indeed a well defined quantity at 
asymptotic times ($\Delta t \rightarrow \infty$) or for very massive 
(classical) particles. 
On the other hand, this relation applies only for a system where particle 
production during the time interval under consideration is negligible.  
In the more general case, we need to assume that particles will be produced 
within the considered time interval. 
We thus obtain
\begin{equation}
	\Delta {N_{ep}} \sim \frac{1}{m \Delta t}
        +\left| \frac{d N_{ep}}{dt} \right| \Delta t,
	\label{u1}
\end{equation}
which, letting $d N_{ep}/dt \sim N_{ep}/\tau_{cl}$, is minimized for 
\begin{equation}
\Delta t = \tau_{mi} = \frac{1}{(m | d N_{ep}/dt |)^{1/2}} 
           \sim 1/(N_{ep} \, e {\cal E})^{1/2}. 
\end{equation}
The particle number is a well defined quantity only if the change in the number
of particles is small within the time we are considering. 
If not, we need to resort to higher order approximations. 
This renormalization technique is referred to as adiabatic regularization 
\cite{p2,p13,p15}. 
We note that $\Delta N_{ep}$ cannot be made arbitrarily small, and it is 
minimized for
$\Delta N_{ep} = (2/m^{1/2}) \left| d N_{ep}/dt \right|^{1/2} 
\sim 2 (N_{ep} e {\cal E})^{1/2}/m$. 
In summary, the quantum Vlasov equation represents a physical observable (the 
number of electron positron pairs) only if the hierarchy of times 
$\tau_{mi} \lesssim \tau_{cl}$ and $\tau_{qu} < \tau_{cl} < \tau_{pl}$ is 
satisfied.

We should note that the NeqQFT framework is not the only one been implemented 
in the calculation of subcritical pair production, and at present a large 
amount of theoretical work has appeared \cite{p17,p18,p19,p20,c2,c3,p21,p22}. 
While the various approaches seem to converge for large electric fields 
\cite{p16}, some discrepancies in the predicted pair number are seen in the 
subcritical regime. 
On the other hand, recent work seems to demonstrate that despite theoretical 
techniques being very different, they are effectively equivalent solution of 
the same problem, with the differences arising only from the details of the 
numerical methods \cite{a1}.  
Still remains the fact that the precise details of the vacuum breakdown 
mechanisms in full spatial and temporal resolution are not yet fully understood
despite the pioneering work of Schwinger \cite{p23}. 
Techniques based on the worldline path integral \cite{a2,a3} as well as 
calculation of the tunneling probabilities of virtual pairs from the Dirac sea 
\cite{a4,a5} have been successful in determining the pair production in 
simplified non-uniform field configurations.
However, experiments in supercritical fields (\ie in the Coulomb field of an 
ion) have shown contradicting results and the question is still open on whether
the Dirac equation is applicable in these scenarios \cite{p22} $-$ see also 
discussions on the Klein paradox \cite{p24} $-$ and the necessity to use a 
multi-body second quantization formalism (as in the NeqQFT approaches) becomes 
clear. 
In simpler terms, as particles are created, their associated electric field 
adds to the external field, which then feeds back to the production of the next
pair.

\section{Solution of the quantum Vlasov equation for idealized fields}
Several attempts have been made to solve the quantum Vlasov equation in 
cosmological regimes \cite{p2,p3,p4}. 
More recently, attention has been drawn to the fact that the new generation of 
laser and FEL facilities has now reached electric field intensities where the 
particle production could have observable effects. 
In the simplest case of a time invariant, spatially homogeneous electric field,
the solution is well known. 
This was originally derived by Schwinger \cite{p2,p23} and found that the pair 
number is exponentially suppressed:
\begin{equation}
	{\cal N}_{\rm Schwinger} 
	\propto \exp\left(-\pi\frac{m^2}{e{\cal E}}\right) 
	= \exp\left(-\pi\frac{{\cal E}_c}{\cal E}\right),
	\label{sw}
\end{equation}
where ${\cal E}_c=m^2/e=1.3\times10^{18}$ V/m is the Schwinger field.
Since the critical field corresponds to the electric field such that its work 
on two electron charges separated by a Compton wavelength equals their rest 
mass, to reach a sizeable rate of pair production we need to have 
${\cal E} \gtrsim {\cal E}_c$. 
On the other hand, the subcritical field regime is defined by the weak field 
condition  ${\cal E} \ll {\cal E}_c$, implying a negligible number of pairs 
being generated.
Equation (\ref{sw}) is only valid for static fields. 
In case of dynamically variable electric fields, the pair production problem 
can be understood as a tunneling with an oscillating barrier. 
This enhances the probability of generation of pairs since the average barrier 
seen by the virtual pair is lower. 

A clear advantage of the quantum Vlasov approach is that it can be used to 
model the full temporal dependence of the particle number for any time 
$t \gtrsim \tau_{cl}$. 
A solution of Equation (\ref{qve}) for a sinusoidal, spatially homogeneous, 
laser field has been recently proposed \cite{p21}. 
In this paper, we will consider instead a different temporal dependence of a 
spatially uniform field at the laser focus of two counter propagating laser 
beams:
\begin{equation}
	{\cal E}(t) = {\cal E}_0 \sinh^2(\nu t),
	\label{e}
\end{equation}
for which the Dirac equation is exactly solvable and analytical approximations 
are easily obtained. 
If we assume that the pair production is modest, \ie $f_{\bf k} \ll 1$, and
${\cal E}_0 \ll {\cal E}_c$, then \cite{a6}
\begin{equation}
	{\cal N}(t) = \frac{1}{2 (2\pi)^3} \int d^3 k 
	\epsilon_{\perp}^2 \left| \int_{-\infty}^t du 
	\frac{e {\cal E}(u)}{\Omega_{\bf k}^2(u)}  
	\exp \left(2 i \int_u^t d\tau \Omega_{\bf k}(\tau)\right)  \right|^2.
\end{equation}
Under the condition $e {\cal E}_0/m \nu \ll 1$, which implies a semi-classical 
motion of the charges in the electric field, the pair number can be further 
simplified to \cite{a6}
\begin{equation}
	{\cal N}(t) \sim \frac{1}{2 (2\pi)^3} \int d^3 k 
	\frac{\epsilon_{\perp}^2}{\Omega_{\bf k}^4(t)} 
	\left| \int_{-\infty}^t du \, e {\cal E}(u) \, 
	{\rm e}^{2 i \Omega_{\bf k} u} \right|^2.
\end{equation}
Using the field (\ref{e}), the asymptotic (residual) pair number density 
becomes
\begin{eqnarray}
	n_r = {\cal N}(t=\infty) = 
	\frac{(e {\cal E}_0)^2}{2 (2\pi)^3} \int d^3 k 
	\frac{\epsilon_{\perp}^2}{\Omega_{\bf k}^4} 
	\left| \frac{2 \pi {\rm csch}(\pi \Omega_{\bf k}/\nu)}{\nu^2} \right|^2
 	\nonumber \\
	\sim \frac{4}{3} \frac{(e {\cal E}_0)^2}{m} 
	\left(\frac{m}{\nu} \right)^4 {\rm e}^{-2 \pi m/\nu},
\end{eqnarray}
where we have assumed $\nu \ll m$, and $\epsilon_{\perp} \sim \Omega_{\bf k}$ 
(which is valid for weak fields). 
Moreover, we have taken $\Omega_{\bf k} \sim m$ for $k\lesssim m$ and 
$\Omega_{\bf k} \sim k$ for $k\gtrsim m$.
We see that in this case, pair production is exponentially suppressed for 
subcritical fields. 
The exponential term ${\rm e}^{-2 \pi m/\nu}$ is indeed equivalent to what 
obtained with other techniques \cite{a4,a5}.
This confirms the fact that, even for oscillating fields, a significant number 
of pairs can persist only for fields close to the Schwinger limit.  
The specific functional form for the residual density is dependent on the 
exact time variation of the electric field, and different results are obtained 
for sinusoidal fields ${\cal E}(t) = {\cal E}_0 \sin(\nu t)$ \cite{p21,a6}. 
In the latter case, the residual density after one oscillation period is 
$n_r \sim (e {\cal E}_0 \nu)^2/m^3$, which is again negligible in the 
subcritical regime.  

While the residual density is exponentially suppressed at asymptotic times for 
the idealized field of Eq. (\ref{e}), the pair density at finite times is 
significantly larger.
The average pair density during the field excitation can be approximated as
\begin{eqnarray}
	n_{av} \sim \frac{{\cal N}(t=0)}{2} = 
	\frac{1}{4 (2\pi)^3} \int d^3 k 
	\frac{\epsilon_{\perp}^2}{\Omega_{\bf k}^4} 
	\left| \int_{-\infty}^0 du \, e {\cal E}(u) \, 
	{\rm e}^{2 i \Omega_{\bf k} u} \right|^2 
\nonumber \\
	\sim \frac{(e {\cal E}_0)^2}{4 (2\pi)^3} \int d^3 k 
	\frac{\epsilon_{\perp}^2}{\Omega_{\bf k}^4} 
	\left| \frac{1}{2 i \Omega_{\bf k}} \right|^2 
\nonumber \\
	\sim \frac{1}{24 \pi^2} \frac{(e {\cal E}_0)^2}{m}.
\end{eqnarray}
Differently from the residual number, the average pair density is not 
exponentially suppressed. 
Moreover, calculation assuming a sinusoidal field showed the same functional 
dependence apart from a numerical prefactor of order unity \cite{p21,a6}. 
This may indicate that the average pair density is not too sensitive on the 
details of the field fluctuations.

Until now we have considered spatially homogeneous fields. 
Real fields, however, are not spatially uniform and variations are expected to 
occur on some macroscopic scale $\Lambda$. 
These effects are more easily estimated within the semiclassical tunneling 
probability calculation \cite{a4,a5}. 
Since in a spatially inhomogenous field pairs are initially produced at the 
maximum of the field, if they move away from this point and the field drops 
too sharply, they may not gain enough energy to cross the barrier and become 
real particles. 
Thus, opposite to the case of time varying fields, spatial gradients tend to 
suppress pair production. 
It can be shown that in the subcritical regime this effect introduces a 
correction to the pair production number of the order \cite{a4,a5}
\begin{equation}
    {\cal C} \sim 1 - \frac{5}{4} \left(\frac{m}{e{\cal E}_0\Lambda}\right)^2,
\end{equation}
where $m/e{\cal E}_0\Lambda \lesssim 1$.

\section{Observable effects from pair production}
As we have discussed in the previous section, in the subcritical regime, pair 
production at asymptotic times is always exponentially suppressed, meaning 
that no residual pairs remains after the laser.
On the other hand, there is a significant number of pairs during the time the 
electric field is switched on. 
Assuming that the laser has wavelength $\lambda$, then the estimated total 
number of electron-positron pairs in the laser spot volume $V \sim s^2 \lambda$
(where $s \gtrsim \lambda$ is the laser spot diameter) is given by 
\begin{equation}
	N_{ep} = V n_{av} {\cal C}
	\simeq \frac{s^2 \lambda}{24 \pi^2} 
	\frac{(e {\cal E}_0)^2}{m} 
	\left[ 1 - \frac{5}{4} \left(\frac{m}{e{\cal E}_0 s}\right)^2 \right],
\end{equation}
where the scale of spatial inhomogeneities is given by the spot size 
($\Lambda \sim s$).

Such number of electron-positron pairs has a clear observable effect, namely 
the generation of gamma rays due to pair annihilation. 
If during the laser pulse the particle number is a well defined physical 
quantity, then collision between those particles are indeed possible.
Since the pair number scales with the laser wavelength (\ie the interaction 
volume at the focal spot), it shows that optical lasers have some advantage 
over x-ray FELs.
On the other hand, for very intense FELs we could have the opposite scenario 
where the spot volume is too small to generate a sizeable number of pairs 
during the evolution of the laser pulse, but the field intensity is large 
enough that a finite number of pairs remains at asymptotic times. 
Those pairs may lead to accumulation effects (\ie interacting with the external
field) and induce spontaneous pair production over several laser cycles 
\cite{p25}, which is a consequence of the non-Markovian character of the 
quantum Vlasov equation.
While the quantum Vlasov equation is a collisionless equation, if collisions 
are a small perturbation, then the momentum distribution of the pairs is 
unaltered and the resultant number of gamma rays is obtained by multiplying 
their distribution functions by the annihilation cross section integrating 
over all the momenta of the pairs \cite{p21}. 
However, since we want to explore the scaling of the laser parameters with the 
observable number of gamma rays, we will follow here an equivalent analytical 
approximation. 
The ratio of electron-positron collisions producing gamma ray annihilation is 
\cite{a7}
\begin{equation}
	R = \sigma_T \left(\frac{{\cal E}_c}{{\cal E}_0}\right) \nu^2,
\end{equation}
where $\sigma_T$ is the Thomson cross section, which applies at low energies 
compared to the rest mass, as in our case.
The model is applicable if collective plasma phenomena take places on a scale 
shorter than the laser cycle, \ie $\tau_{pl}<2\pi/\nu$, thus allowing 
sufficient time for the particle to interact \cite{p13,p29}.
This also shows that as the laser field decreases, annihilation processes 
becomes more probable. 
Physically, this means that pairs produced with smaller momenta are more likely
to result in a collision event.
The number of gamma rays emitted is thus
\begin{eqnarray}
	\tilde{N}^{\gamma \gamma} \simeq N_{ep} R = 
	\frac{ \sigma_T s^2 \lambda \nu^2}{24 \pi^2} 
	\left(\frac{e^2 {\cal E}_c {\cal E}_0}{m}\right)
	\left[1 - \frac{5}{4} \left(\frac{m}{e{\cal E}_0 s}\right)^2 \right] 
\nonumber \\
	\simeq \frac{\sigma_T s^2}{3 \lambda} 
	\left(\frac{e^2 {\cal E}_c {\cal E}_0}{m}\right) 
	\left[ 1 - \frac{5}{4} \left(\frac{m}{e{\cal E}_0 s}\right)^2 \right],
\end{eqnarray}
where we have used the fact that $\nu \sim 2 \pi /\lambda$ (which is exact for 
a sinusoidal wave). 
While the treatment presented here is far from being complete, the values 
obtained with this approach are in agreement with the predicted number of 
gamma rays calculated by full integration of the pair distribution function 
from the quantum Vlasov equation \cite{p21}.
Both approaches are, however, not self consistent, and a complete analysis will
require the addition of a collisional sink term in the quantum Vlasov equation
\cite{p14,p26}, which must then be coupled to the gamma ray production rate. 
Only in this way the full effects of entanglement and quantum statistics could 
be properly accounted for. 

In order to get a realistic value for the expected number of  gamma rays, we 
also need to account for the fact that counter propagating beam geometries are 
experimentally difficult to realize (see, however Ref.~\cite{p27} for a 
suggested counter-propagating beam geometry). 
If $\theta$ is the angle between the two beams, then this introduces a 
geometrical correction $(1-\cos\theta)/2$. 
Moreover, if the laser beam has a pulse duration $\tau_L$, then gamma ray 
annihilation events will occur $\tau_L \nu/ 2 \pi$ times during the laser shot.
 
Bringing back the factors of $c$ and $\hbar$, the total
number of expected $\gamma \gamma$ events during a laser pulse is then
\begin{equation}
	N^{\gamma \gamma} 
	\simeq \frac{(1-\cos\theta) \, \sigma_T \, m c^3 \, s^2\, \tau_L \, 
	e {\cal E}_0}{3 (\hbar c \lambda)^2} 
	\left[1-\frac{5}{4}\left(\frac{m c^2}{e{\cal E}_0 s}\right)^2 \right],
	\label{ng}
\end{equation} 
with the electric field is expressed in terms of the laser intensity, $I_0$, 
as ${\cal E}_0 = (2 \mu_0 c I_0)^{1/2}$ (in SI units).

\section{Photometric}
In this section we compare expected number of $\gamma$ photons from pair 
annihilation with respect to background noises.
As shown in Table \ref{t1}, we expect $\sim$0.6 annihilation events per laser 
shot, corresponding to $\sim$10000-events in a 10 hr experiment using the 
Astra Gemini laser available at the Rutherford Appleton Laboratory. 
In a full experimental week, this corresponds to $5\times10^4$ annihilation 
events producing two gamma ray photons. Coincidence measurements will be 
performed with high sensitivity large area NaI gamma ray detectors covering a 
solid angle of $\sim 2 \pi$, with an absolute conversion efficiency $>0.08$ 
\cite{p28}. 
We can estimate a total detection of $\sim 2\times10^3$ events. 
In situ measurements to assess the background level within the laser area have 
observed 2060 positron events in 10 hours, equivalent to 0.05 counts/s. 
Since the NaI detectors can be gated with integration time $\sim$1 $\mu$s, the 
background level of cosmic ray hits can be minimized to $\sim$0. 
Any sporadic background event could be further eliminated with a coupled 
anti-coincidence detector. 

The major source of noise in these experiments arises from bremsstrahlung 
photons emitted by electrons stripped from the residual gas in the laser
focal spot. Since relativistic electrons will be produced at laser intensities 
$I_0 \gtrsim 10^{19}$ W/cm${}^2$, this corresponds to a much larger volume 
than the laser focal spot.
For the Astra Gemini laser, at pressures $\sim 10^{-6}$ mbar, we expect up to 
$10^4$ electrons being ejected by the residual atoms (mostly hydrocarbons and 
oxygen).
If these electrons are all emitted in a narrow cone, the probability that each 
one of them collides with a residual atom before reaching the chamber walls 
($\sim$1 m path length) is less than $10^{-4}$. 
During such a collision a gamma ray photon is emitted, corresponding to 
$< 0.04$ events detected per laser shot (0.002 counts/s).
If the gamma detectors are all placed within 1 m from the laser interaction 
point and outside the stripped electrons path, no additional gamma ray event 
will be recorded, as electrons hitting the chamber walls will emit photons in 
the forward direction away from the detector units (as well as excluded by 
coincidence detection). 

We notice from Table 1 that the error in the number of pairs is substantially 
larger than 1. 
This implies that the (rest) energy of those pairs is undetermined with an 
error (for the Gemini laser) of 1.3 MeV. 
Alternatively, this results can be interpreted in the sense that only a 
fraction of pairs has materialized on the mass shell, but the rest are still 
virtual. 
However, since the detection efficiency of scintillators remains the same over
the $\sim$1 MeV range (centered at 0.511 MeV) \cite{det1}, we would expect that
at worst a count rate is reduced by a factor of 2, to 0.025 counts/s 
(1030 positron events in 10 hours), but still significantly above background.

\begin{center}
\begin{table}
\caption{Operation parameters for current laser systems and expected 
$\gamma \gamma$ yield. 
A beam crossing angle of $\theta=135^{\rm o}$ has been assumed.
The Astra Gemini and the Vulcan PW systems are both located at the Rutherford 
Appleton Laboratory.}
\vspace{5mm}

\begin{tabular}{|c|c|c|}
\hline
		& {\bf Astra Gemini} 			& {\bf Vulcan PW}\\
\hline
Wavelength (nm)		& 800				& 1064  	\\
Pulse length (fs)	& 30		 		& 500		\\
Laser energy (J)	& 15				& 500 		\\
Spot diameter ($\mu$m)	& 5				& 5		\\
Intensity (W/cm${}^2$)	& $2.5\times10^{21}$		& $5\times10^{21}$\\
${\cal E}_0$ (V/m)	& $1.4\times10^{14}$		& $1.9\times10^{14}$\\
$n_{av}$ (cm${}^{-3}$)	& $8.0\times10^{20}$		& $1.6\times10^{21}$\\
$N_{ep}$		& $1.6\times10^{10}$		& $4.2\times10^{10}$ \\
$\tau_{mi}$ (fs)	& $9.9\times10^{-10}$		& $5.1\times10^{-10}$\\
$\tau_{qu}$ (fs)	& $8.1\times10^{-6}$	 	& $8.1\times10^{-6}$ \\
$\tau_{cl}$ (fs)	& $1.2\times10^{-2}$		& $8.7\times10^{-3}$ \\
$\Delta N_{ep}$		& $2.6\times10^3$		& $5.0\times10^3$ \\
$2 \pi \tau_{pl}/\nu$	& 0.13 				& 0.22 \\
$N^{\gamma \gamma}$	& 0.63				& 0.21 	\\
Repetition rate		& every 20 sec			& every 1 hr\\
$N^{\gamma \gamma}$ after 10 hr	& 10879			& 805	\\
\hline
\end{tabular}
\label{t1}
\end{table}
\end{center}

\section{Conclusions}
We have presented a proposal to test subcritical pair production with high 
intensity lasers.
Using the theoretical framework of NeqQFT we have shown that the residual pair 
density after the laser shot is exponentially suppressed, and the number of 
pairs remaining is negligible. 
However, for realistic laser conditions, there is a significant number of pairs
during the field evolution and the observable effect of such pairs is the 
production of co-incident gamma rays.
We have estimated for the Astra Gemini laser facility at the Rutherford 
Appleton Laboratory more than $10^4$ annihilation events during an 
experimental day.
Photometric analysis has shown that this number of events will be detectable 
with current instrumentation.
We are proposing an experimental platform that could test, for the first time, 
NeqQFT models which are relevant to astrophysical and cosmological processes,
and, at the same time, resolve issues with the current approximation schemes of
non-perturbative QED.

This work was supported in part by the Science and Technology Facilities 
Council of the United Kingdom, by Department of Energy, Office of Nuclear 
Physics, contract no. DE-AC02-06CH11357 and by the Polish Ministry of Science 
and Higher Education under grant no. N N 202 0953 33.
D.B.B., A.V.P., G.R. and S.A.S. are grateful for support from the Helmholtz 
Association for their participation at the Summer School on {\it Dense Matter 
in Heavy-Ion Collisions and Astrophysics} in Dubna, July 14-26, 2008, where 
this project has been started. 
D.B.B. thanks D. Habs and G. Mourou for enlightening discussions and 
encouragement during the ELI workshops, in particular the one at Frauenw\"orth,
October 2008.
G.G. would also like to thank T.~Heinzl for useful discussions about the 
manuscript.

\end{document}